\documentclass[shortnote,twocolumn]{jpsj2}
\usepackage{graphicx}

\def\runtitle{loc2002}
\def\runauthor{Masato \textsc{Kishi} and Yasuhiro \textsc{Hatsugai}}

\title{%
Anderson Localization and Polarization
}

\author{%
Masato \textsc{Kishi}$^{1}$\thanks{E-mail: kishi@pothos.t.u-tokyo.ac.jp} and
Yasuhiro \textsc{Hatsugai}$^{1,2}$
}

\inst{%
$^1$Department of Applied Physics, University of Tokyo, 7-3-1 Hongo\\
Bunkyo-ku, Tokyo 113-8656, Japan \\
$^2$PRESTO, JST
}

\recdate{\today}

\abst
{
}

\kword{%
Anderson localization, polarization, dielectric response, density
matrix renormalization group, exact diagonalization
}

\begin{document}
\sloppy
\maketitle

\section{Introduction}
Effects of randomness have supplied fundamental problems
in condensed matter physics and localization due to
interference of quantum mechanical electrons are
well studied as the Anderson localization.
Although we have well established understanding
of the localization of non-interacting electrons,
information of the correlated electrons with randomness is
still missing. It was mainly due to lack of reliable numerical
techniques for the correlated electrons.

Today, for the one dimensional correlated systems without randomness,
lots of numerical results are collected by the
Density Matrix Renormalization Group (DMRG)\cite{rf:1} method
and consistent understanding with analytical predictions
has been achieved.  In this paper, we plan to apply DMRG
for the random electron systems by calculating direct
responses of the system with electric field\cite{rf:4}.
At first,  random systems without interaction are carefully investigated.
Then we try to treat both of
  interaction and randomness in one dimensional systems.

\section{Model and Dielectric response}
We investigate a one-dimensional spinless fermion model with
nearest neighbor electron-electron interaction in the presence of random potential. 
The Hamiltonian is given as
\begin{equation}
H=-t\sum_{i=1}^{L-1} (c_i^\dagger c_{i+1} + c_{i+1}^\dagger c_i) + V
\sum_{i=1}^{L-1} n_in_{i+1} + \sum_{i=1}^L \epsilon_i c_i^\dagger c_i,
\end{equation} 
where $n_i = c_i^\dagger c_i$ and $\epsilon_i$ is random potential,
which distributes over the interval $[-W/2,W/2]$ uniformly. 
We set $t = 1$ for simplicity and consider the half-filling case
and impose an open boundary condition. 
In the absence of disorder, the system is metallic for $-2<V<2$. But for
half-filling case, at $V=2$ the system undergoes a metal-insulator
transition and for $V>2$ the system has a finite charge gap. In
attractive interaction region at $V=-2$, the system becomes unstable
due to phase separation. 
When the randomness is present, the system is always insulating due to
the Anderson localization without interaction. The interplay between the
randomness and interaction can be interesting and possible metallic
phase is expected for some range of negative $V$.$^{3-5)}$

In this parer,
we focus on the dielectric response of the system. 
In order to observe the dielectric response, we apply the electric field $E$ to
the system. As the second quantized form of the potential, $-Ex$, the coupling term $H_E$ is added to the Hamiltonian. Then full
Hamiltonian of the system is given by $H_T = H + H_E$ where
\begin{equation}
 H_E = - E \sum_{i = 1}^L\Big(i - \frac{L}{2}\Big) n_i.
\end{equation}

As a function of $E$, the polarization $P$ of
 the system is defined by
\begin{equation}
 P =- \frac{1}{L} \frac{ \partial E_{0}}{\partial E }
=- \frac{1}{L} \frac{ \partial}{\partial E}\langle H_T \rangle _E
= \frac{1}{L} \sum_{i = 1}^L \Big(i-\frac{L}{2}\Big) \langle n_i  \rangle_E
\end{equation}
where $E_0$ is a ground state energy and $\langle n_i  \rangle _E$
represents the ground state expectation value of $n_i$\cite{rf:4}.
Here we used the Feynman's theorem to derive the last equation.

For a finite value of the electric field, which is comparable to the Mott gap,
we expect a collapse of the local charge gap due to the
interaction. 
Also in the Anderson insulator, reconstruction of the charge by
transferring electrons above the tunneling barrier could occur. In each
case, we can obtain information of the charge degree of freedom above
its ground state.

We turn next to the linear response regime. In this regime, we calculate zero-field dielectric susceptibility as
\begin{equation}
\chi = \left. \frac{\partial P}{ \partial E} \right|_{E = 0}
= - \left. \frac{1}{L} \frac{\partial^2 E_0}{\partial E^2} \right|_{E=0}.
\end{equation}

From the susceptibility $\chi$, we directly obtain information whether
the ground state is metallic or an insulator.
In thermodynamic limit, $\chi$ is diverging if the system is metallic, but
converging to a finite value if it is an insulator.
Indeed $\chi \sim L^2$ is expected by the perturbation for the pure
non-interacting system($W=0,V=0$).

In order to calculate the charge distribution and the ground state
energy, we use the exact diagonalization for $V=0$, and DMRG
for finite $V$.
For the application of DMRG, we use the extended infinite-size
algorithm by Hida\cite{rf:3}, which enables us to treat also
non-uniform lattice models. We perform
three or four finite lattice sweep for the convergence.
The retained
states for the block is 60-100 to keep the truncation error to be less
than $10^{-8}$. 

\section{Results and Discussion}
\begin{center}
  \begin{figure}
\hspace*{0.33cm}
   \includegraphics[scale=0.51]{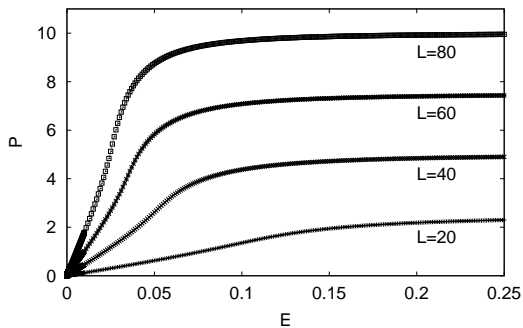}
    \caption{Polarization $P$ as a function of applied electric field
$E$. ($W=0$ and $L =$ 20, 40, 60 and 80.)}
    \label{fig1}
  \end{figure}
\end{center}

\begin{center}
  \begin{figure}
\hspace*{0.19cm}
   \includegraphics[scale=0.51]{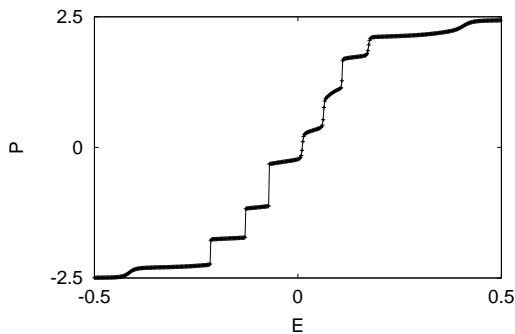}
    \caption{Polarization $P$ as a function of applied electric field
$E$. ($W = 5.0$ and $L = 20$.)}
    \label{fig2}
  \end{figure}
\end{center}

At first using exact diagonalization, we calculate the polarization
$P$ as a
function of the applied electric field $E$ in the absence of electron-electron
interaction. The typical $P-E$ curve is shown in Fig. \ref{fig1} and
Fig. \ref{fig2}.~The polarization $P$ is a smooth function of $E$ without randomness. 
As we increase the electric field from zero, $P$ is almost linearly
increasing and approaches to a finite value, at which all electrons are
collected to one side. Then at the critical field $E_s \sim 1/L$, $P$
saturates as $P \sim~L$.

In Fig. \ref{fig2}, we plot the $P-E$ curve for finite randomness
strength $W$.
In the presence of randomness, the $P-E$ curve
exhibits a stepwise behavior.
This step is caused by crossing of the one particle energy levels. 
Namely, at some critical $E$, level crossing between
the highest occupied state and the lowest unoccupied state occurs.
By this process, the charge reconstruction of the ground state occurs
which corresponds to the electron tunneling between localized states.


\begin{center}
  \begin{figure}[h]
\hspace*{1.0cm}
   \includegraphics[scale=0.50]{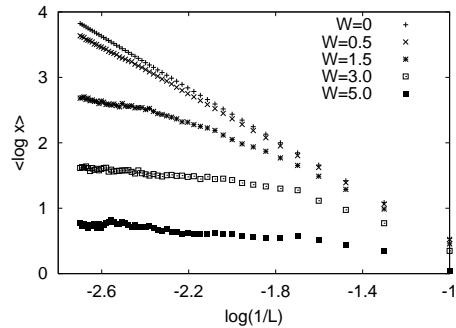}
    \caption{Average $\log\chi$ as a function of $\log1/L$.
    The system sizes are between 10 and 500. ($V=0$. $W =$ 0, 0.5,
1.5, 3.0 and 5.0.)  }
\label{fig3}
  \end{figure}
\end{center}

Next let us focus on the linear response region. In Fig.~\ref{fig3}, 
we show the susceptibility $\chi$ as a function of $1/L$ by
 numerically differentiating the ground state energy. We averaged over
500 realizations of the
disorder potential and we applied the electric
field $E =\pm 10^{-5}$. For $W=0$, $\chi$ increases as $\chi \sim L^2$ and
$\chi$ seems to diverge. This means the system is metallic. On the
other hand, for $W=3$, $\chi$ is convergent to a finite value
which is consistent with the insulating ground state.
 For intermediate $W$, for example $W=0.5$,
saturation of $\chi$ is not observed up to $L = 500$. The localization
length $\xi$ seems to be larger than the system length we used. Then finite size
scaling analysis is necessary. The detail can be published elsewhere.

\begin{center}
  \begin{figure}[h]
\hspace*{1.0cm}
   \includegraphics[scale=0.46]{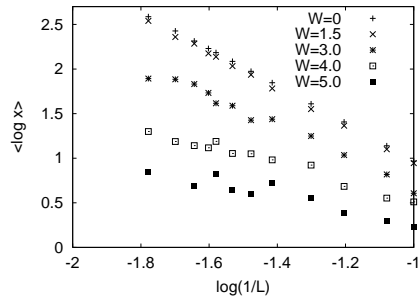}
    \caption{Average $\log \chi$ versus $\log 1/L$ for $V = -1.4$. The
system sizes are 10, 12, 16, 20, 26, 30, 34, 38, 40, 44, 50 and 60.
($W=$ 0, 1.5, 3.0, 4.0 and 5.0.) }
\label{fig4}
  \end{figure}
\end{center}

Using DMRG, we study also the systems with electron correlation. 
In Fig. \ref{fig4}, $\log\chi$ versus $\log 1/L$ for $V = -1.4$ is plotted.
We averaged over 64 realizations of the disorder potential.
When the randomness is sufficiently strong, we observe the
saturation, which implies the localized ground state. However, we need more extensive analysis to obtain
conclusive results.
The detail analysis with finite size scaling will be given elsewhere. 

The computation in this work has been done in part using the
facilities of the Supercomputer Center, ISSP, University of Tokyo.


\begin{thebibliography}{99}
\bibitem{rf:1} S. R. White: 
Phys. Rev. Lett. {\bf 69} (1992) 2863; 
Phys. Rev. {\bf B 48} (1993) 10345.
\bibitem{rf:4} C. Aebischer, D. Baeriswyl and R. M. Noack: 
Phys. Rev. Lett. {\bf 86} (2001) 468. 
\bibitem{rf:6} T. Giamarchi and H. J. Schulz:
Phys. Rev. {\bf B 37} (1988) 325.
\bibitem{rf:2} G. Bouzerar and D. Poilblanc:
J. Phys. I (France) {\bf 4} (1994) 1699.
\bibitem{rf:5} P. Schmitteckert {\it et al.}:
 Phys. Rev. Lett. {\bf 80} (1998) 560.
\bibitem{rf:3} K. Hida: 
J. Phys. Soc. Jpn. {\bf 65} (1996) 895.
\end{thebibliography}
\end{document}